\documentclass[showpacs]{revtex4}
\usepackage{amsmath}
\usepackage{amssymb}
\usepackage{epsfig}

\newcommand{\be}{\begin{eqnarray}}
\newcommand{\ee}{\end{eqnarray}}

\def\bra#1{\textstyle{\left\langle \, #1 \, \right\vert \>}}
\def\ket#1{\textstyle{\> \left\vert \>\! #1 \>\! \right\rangle}}

\begin{document} 
%\title{\hfill {\footnotesize FZJ--IKP(TH)--2005--21} \\
\title{
Extraction of scattering lengths from 
final-state interactions}

\author{A. Gasparyan$^1$, J. Haidenbauer$^2$, and C. Hanhart$^2$}

\affiliation{
$^1$Institute of Theoretical and Experimental Physics,
117259, B. Cheremushkinskaya 25, Moscow, Russia\\
$^2$Institut f\"ur Kernphysik (Theorie), Forschungszentrum J\"ulich,
D-52425 J\"ulich, Germany
}

\begin{abstract}
A recently proposed method based on dispersion theory, that allows to extract 
the scattering length of a hadronic two-body system from corresponding
final-state interactions, is generalized
to the situation where the Coulomb interaction is present. 
The steps required in a concrete practical application are discussed in detail. 
In addition a thorough examination of the accuracy of the proposed method
is presented and a comparison is made with results achieved with other    
methods like the Jost-function approach based on the effective-range 
approximation. Deficiencies of the latter method are pointed out. 
The reliability of the dispersion theory method for extracting also the 
effective range is investigated. 
\end{abstract}

\maketitle

\section{Introduction}

The scattering length 
provides not only an important measure for the strength of the 
interaction in a specific hadronic two-body system \cite{Joach}
but often allows to draw further more general and thus even more
interesting conclusions. For example in the case of the
proton-proton and neutron-neutron systems the corresponding
scattering lengths in the $^1S_0$ partial wave provide a 
very sensitive test of charge symmetry in the strong interaction \cite{Miller}.
SU(3) symmetry can be tested by comparing the $^1S_0$ scattering
lengths in the neutron-proton and $\Sigma^+p$ systems, which should
fulfil the relation $a_{np} = a_{\Sigma^+p}$ in case SU(3) symmetry
holds rigorously \cite{Dover}. 
In the chiral limit the $\pi N$ S-wave scattering
lengths vanish  and therefore any deviation from that value
is a direct measure for how strongly this symmetry is broken. Here especially
the isoscalar component is of high interest due to its close link to the sigma
term of the nucleon \cite{xx}. 
On a more phenomenological level the $\eta{N}$ scattering length is
interesting because the magnitude of its real part is directly
linked with speculations about the existence of $\eta$-mesonic
hadronic bound states such as $\eta ^3He$
\cite{Wycech,Belyaev1,Belyaev2,Rakityansky,Fix1,Niskanen,Niskanen1} 
or $\eta ^4He$ \cite{Haider2}. 

Unfortunately, a direct determination of the scattering length is
only feasible in a few cases. It can be done with 
scattering experiments sufficiently close to the reaction threshold
so that the effective range expansion can be utilized for 
extracting the scattering length. But in practice such experiments are
only possible for charged and also (quasi) stable particles,
as it is the case with, e.g., $pp$, $\pi^+p$ or $K^+p$ scattering. 
One can also extract the scattering length from a study of the hadronic 
level shifts of atoms \cite{rus} like $\pi^-p$,
$\pi^-d$ or also $\bar p p$ \cite{Gotta}. 
For the majority of the hadronic two-body systems
information about the scattering length is only accessible
via an investigation of the final-state interaction of systems
which have at least three particles in the final state.  
 
While in the former type of experiments the accuracy of the scattering 
length is directly connected with the precision of
the data more detailed and often sophisticated considerations are necessary 
in order to estimate additional uncertainies that arise when the scattering 
length is extracted from final-state effects \cite{Gibbs}. However, in some 
cases such an
estimation is facilitated by the fact that the reaction mechanism is known. 
For example, the reaction $nd \to (nn)p$, one of the prime sources of the
$nn$ scattering length, can be analysed by means of rigorous Faddeev
calculations \cite{Trotter,Huhn,Deng}. Reactions involving the pion such as 
$\pi d$, $\pi^- d\to \gamma nn$ or $\gamma d \to \pi^+ nn$, which can be used to extract the $\pi N$ and
$nn$ scattering lengths, respectively, can be tackled by chiral perturbation
theory in a well controlled way in the relevant near-threshold regime \cite{xxx,Lensky}. 

In a recent publication \cite{Gasparyan2003} 
we argued that also large-momentum transfer reactions such as
$pp\to K^+p\Lambda$ \cite{C11_1,Bilger,jan} or $\gamma d\to K^+n\Lambda$
\cite{Renard,Kerbikov,Mecking,Adel,Li,Yama} are excellent candidates for extracting
information about the scattering lengths. In reactions with large
momentum transfer the production process is necessarily of short-ranged
nature. As a consequence the results are basically insensitive to details of
the production mechanism and therefore a reliable and general error estimation can
be given. 
Indeed, in \cite{Gasparyan2003} a formalism based on dispersion theory was 
presented that relates spectra from large-momentum transfer reactions,
such as $pp\to K^+p\Lambda$ or $\gamma d\to K^+n\Lambda$, directly to
the scattering length of the interaction of the final state particles.
The theoretical error of the method was estimated to be 0.3 fm or
even less, which is comparable to the error quoted in the context of the
determination of $a_{nn}$ \cite{Howell}, say.
This estimate was confirmed by comparing results obtained with
the proposed formalism to those of microscopic model calculations for the
specific reaction $pp\to K^+\Lambda p$.
The arguments but also the formalism of Ref. \cite{Gasparyan2003} are, of course, 
valid for any production or decay process which is of short-ranged nature, 
i.e. also for investigation of hadronic two-particle subsystems resulting from the 
decay of the $J/\Psi$ or $B$ mesons \cite{Psi,Belle}. 

In the present paper we want to investigate further aspects of extracting scattering
lengths from final-state interactions which were not addressed in our
earlier work. One of those topics is the presence of the Coulomb interaction. 
In many interesting hadronic two-particle systems both particles carry 
charges, like in the already mentioned $\Sigma^+p$ channel whose scattering
length could be extracted from the reaction $pp\to K^0 \Sigma^+p$.
Then the production amplitude acquires additional singularities, due to the
long-range nature of the Coulomb forces, and the formalism developed
in Ref. \cite{Gasparyan2003} is no longer directly applicable. 
We will derive the modifications that are necessary in order to adapt the
dispersion-relation method to the situation when the Coulomb force
is present in the final-state interaction. We also demonstrate in a toy model
calculation how one has to proceed in a concrete application to data. 

In addition we present a more detailed examination of the accuracy of the 
method proposed in Ref. \cite{Gasparyan2003}.
A test based on one specific model calculation, namely for the
reaction $pp\to K^+\Lambda p$, has been already performed in that paper. 
However, here we want to put this investigation on a broader basis by considering
final-state interactions of varying strengths, corresponding to a much larger range
of values of the scattering length. In addition we take a look at the effective
range $r_e$ as well which can be also extracted by the proposed dispersion-integral
method. For the effective range a sensible error estimation is not possible, as was
already pointed out in Ref. \cite{Gasparyan2003}, but it is still interesting to
examine in concrete applications whether meaningful results could be achieved. 
Finally, and equally important, we want to
compare the present method with the performance of other, approximative treatments
of the final-state interaction that are commonly used in the literature to
extract information on the scattering length and also the effective range.
This concerns in particular the Jost-function approach \cite{book} based on the 
effective-range approximation 
and an even simpler approach that relies simply on utilizing the effective range 
approximation itself \cite{migdal}. Thereby, we will show that the latter methods lead to 
(partly drastic) systematic deviations
from the true values and therefore one has to be rather cautious in the 
interpretation of results achieved with those methods. 

The paper is structured in the following way: In the subsequent section we
give a short review of the dispersion integral method for extracting the scattering
length from final-state interactions. In section 3 results of an examination of the 
accuracy of this method are presented. Thereby,
we consider various (singlet and triplet) $S$ wave $YN$ and $NN$ interactions
and compare the scattering lengths and effective range extracted with the 
dispersion-integral method from appropriately generated final-state effects 
with the ones predicted by the models. We also apply two approximative methods
for treating final-state effects, namely the Jost-function approach
based on the effective range approximation (Jost-ERA) 
as well as the effective range approximation itself, 
and compare their performance with the one of our method. 
In section 4 we generalize the dispersion-integral method to the case where
a repulsive Coulomb interaction is present in the final state. Test calculations for
a final-state interaction with Coulomb are then presented in sect. 5 and it
is discussed in detail how one has to procede in a practical application. 
The paper ends with a short summary.

\section{Formalism}
Our method, which goes back to an idea of Geshkenbein 
\cite{Geshkenbein1969,Geshkenbein1998}, 
is based on using the dispersion relation technique.
Consider the production amplitude $A_S$ of a $2\to 3$ reaction. To be
concrete we discuss $pp\to
K^+p\Lambda$, or $\gamma d\to K^+n\Lambda$, with the 
$\Lambda N$ system being in an $L=0$ partial wave and a specific spin state 
$S$ ($^1S_0$ or $^3S_1$). This amplitude depends on the total energy 
squared $s=(p_1+p_2)^2$, the invariant mass squared of the
outgoing $\Lambda N$ system $m^2=(p_N+p_\Lambda)^2$ and the
momentum transfer $t=(p_1-p_{K^+})^2$, where $p_1$, $p_2$, $p_N$, $p_\Lambda$,
and $p_{K^+}$ are the 4-momenta of the two initial particles, final nucleon,
lambda, and kaon, respectively. Then one can write down a dispersion relation
for this amplitude with respect to $m^2$ at fixed $s$ and $t$
\begin{eqnarray}
A_S(s,t,m^2)=\frac1\pi\int_{-\infty}^{\tilde m\, ^2} \frac{D_S(s,t,m' \, ^2)}{m' \, ^2-m^2}dm' \, ^2
+\frac1\pi\int_{m_0^2}^\infty \frac{D_S(s,t,m' \, ^2)}{m' \, ^2-m^2}dm' \, ^2,
\label{dispers}
\end{eqnarray}
where $\tilde m\, ^2$ is the upper boundary of the lefthand cut, $m_0^2 = (m_N+m_\Lambda)^2$, and
\begin{eqnarray}
D_S(s,t,m^2) = \frac{1}{2i}(A_S(s,t,m^2+i0)-A_S(s,t,m^2-i0))
\end{eqnarray}
is the discontinuity of the amplitude along the cuts.
We neglect here the contributions from possible kaon-baryon interactions.
In case they are not small, they still can be considered as constant (weakly
mass dependent) if one chooses the kinematics such that the excess energy of
the reaction is significantly larger than the typical range of the $\Lambda N$
interaction, cf. the discussion in Ref. \cite{Gasparyan2003}.
The index $S$ denoting the spin state will be suppressed in the following to
simplify the notation. 

For a purely elastic $\Lambda N$ system, the discontinuity along the righthand 
cut would be given by 

\begin{eqnarray}
D(s,t,m^2) = A(s,t,m^2)e^{-i\delta}\sin{\delta},
\label{drhc}
\end{eqnarray}
where $\delta$ is the  $\Lambda N$ ($^1S_0$ or $^3S_1$) scattering phase
shift.
Then the solution of Eq. \eqref{dispers} in the physical region reads 
(see Refs. \cite{Muskhelishvili1953,Omnes1958,Frazer1959})
\begin{eqnarray}
A(s,t,m^2)=\exp\left[{\frac1\pi\int_{m_0^2}^\infty\frac{\delta(m' \, ^2)}{m' \, ^2-m^2-i0}dm' \, ^2}\right]
\Phi(s,t,m^2),
\label{dis0}
\end{eqnarray}
where
$\Phi(s,t,m^2)$ contains only lefthand singularities and therefore is  a
slowly varying function of $m^2$. In order to ensure this requirement to be
fulfilled it is important that the momentum transfer $t$ is large.
We assume also that there is no bound state in the  $\Lambda N$ system.

Consider now a realistic situation where inelastic channels are present -- as it is the
case with $\Lambda N$ due to the coupling to the $\Sigma N$ channel, say.
Then one can write down a formula similar to Eq.~\eqref{dispers}, but with
the integration performed over a finite range of masses \cite{Gasparyan2003}:
\begin{eqnarray}
A(m^2)=\exp\left[{\frac1\pi\int_{m_0^2}^{m_{max}^2}\frac{\delta(m' \, ^2)}{m' \, ^2-m^2-i0}dm' \, ^2}\right]
\tilde \Phi(m^2) \label{C_int} \ ,
\label{A_def} 
\end{eqnarray}
where $\tilde \Phi(m^2)$ is again a slowly varying function of $m^2$ given
the phase shift $\delta$ is sufficiently small in the vicinity of $m_{max}$ \cite{Gasparyan2003}.
The upper limit $m_{max}$ has to be chosen in such a way that the corresponding relative 
momentum of the $\Lambda N$ system, $p_{max}$, is of the order of the typical
scale of the $\Lambda N$ interaction, i.e. in the order of $1/a$ or $1/r$.
The equation \eqref{A_def} can be solved with respect to the $\delta$ \cite{Gasparyan2003}:
\begin{eqnarray}
\nonumber
&&\frac{\delta(m^2)}{\sqrt{m^2-m_0^2}}= \\
& & -\frac1{2\pi}{\bf P}
\int_{m_0^2}^{m_{max}^2}\frac{\log{|A(m' \, ^2)/\tilde \Phi(m_{max}^2,m' \, ^2)|^2}}
{\sqrt{m' \, ^2-m_0^2} \ (m' \, ^2-m^2)}\sqrt{\frac{m_{max}^2-m^2}{m_{max}^2-m' \, ^2}}dm' \, ^2.
\label{almostfinal}
\end{eqnarray}
If one neglects the mass dependence of  $\tilde \Phi(m^2)$  and uses the
relation between the partial cross section $\sigma_S$ and the amplitude
$$\frac{d^2\sigma_S}{dm' \, ^2dt} \propto p'|A_S(s,t,m'\,^2)|^2 \ ,$$ 
then one obtains the expression for the scattering length in terms
of observables
\begin{eqnarray}
\nonumber
a_S&=&\lim_{{m}^2\to m_0^2}\frac1{2\pi}\left(\frac{m_\Lambda+m_N}
{\sqrt{m_\Lambda m_N}}\right){\bf P}
\int_{m_0^2}^{m_{max}^2}dm' \, ^2
\sqrt{\frac{m_{max}^2-{m}^2}{m_{max}^2-m' \, ^2}}\\
& & \qquad \qquad \times \ 
\frac1{\sqrt{m' \, ^2-m_0^2} \ (m' \, ^2-{m}^2)}
\log{\left\{\frac{1}{p'}\left(\frac{d^2\sigma_S}{dm' \, ^2dt}\right)\right\}}
\ ,
\label{final}
\end{eqnarray}
and analogously for the effective range $r_e$.

\section{Accuracy of the method and comparison with other approaches}

The most important advantage of the method proposed by us \cite{Gasparyan2003}
is that a reliable estimate for the uncertainty of the extracted scattering
length can be given. The sources for the uncertainty are trifold: (i) A possible
influence of the final-state interaction in the other outgoing channels. For the
reaction $pp\to K^+\Lambda p$ considered in Ref.~\cite{Gasparyan2003} this
concerns the $K\Lambda$ and $KN$ systems.  (ii) The adopted value for $m^2_{max}$, the 
upper limit chosen for the dispersion integral in Eq. (\ref{final}). 
(iii) A sensitivity to left-hand cuts of the production operation. 
A detailed analysis of the issues (ii) and (iii) , based on general arguments, 
presented in Ref.~\cite{Gasparyan2003} suggests that the error in the scattering 
length should be typically in the order of 0.3 fm or less. The role of issue (i) 
cannot be quanitified theoretically but has to be investigated by performing experiments 
and corresponding analyses at different beam momenta \cite{Gasparyan2003}. 

In this section we want to present a thorough examination of the accuracy of the proposed
method and, in particular, to corroborate the error estimate, by means of concrete
model calculations. A test based on one specific model calculation, namely for the 
reaction $pp\to K^+\Lambda p$, has been already performed in Ref.~\cite{Gasparyan2003}. 
However, here we want to put this investigation on a broader basis by considering 
final-state interactions of varying strengths, corresponding to a much larger range
of values of the scattering length. Furthermore, and equally important, we want to
compare the present method with the performance of other, approximative treatments
of the final-state interaction that are commonly used in the literature to 
extract information on the scattering length and the effective range 
and have been applied to $pp\to K^+\Lambda p$ \cite{Bale,Hinter}.

%%%
One of those approximative treatments follows from the assumption that the 
phase shifts are given by the first two terms in the effective range
expansion, 
\begin{eqnarray}
p \ {\rm cot} (\delta(m^2)) = -{1 \over a}+{r_e\over 2} p \,^2 \ ,
\label{ere}
\end{eqnarray}
usually called the effective range approximation (ERA), over the whole energy 
range. Here $p$ is the
relative momentum of the final-state particles under consideration
in their center of mass system, corresponding to the invariant mass $m^2$.
In this case the relevant integrals \eqref{dis0} can be evaluated in closed
form as \cite{book}
\begin{eqnarray}
A(m^2) \propto 
\frac{(p^2+\alpha^2)r_e/2}{-1/a+(r_e/2)p^2-ip} \ ,
\label{arform}
\end{eqnarray}
where $\alpha = 1/r_e(1+\sqrt{1-2r_e/a})$. 
Because of its simplicity Eq. (\ref{arform}) is often used for the treatment 
of the final-state interaction (FSI).

A further simplification can be made if one assumes that 
$a\gg r_e$. This situation is practically realized in the $^1S_0$ partial
wave of the $NN$ system. Then 
the energy dependence of the quantity in Eq. (\ref{arform})
is given by  the energy dependence of the elastic  amplitude
\begin{eqnarray}
A(m^2) \propto 
\frac{1}{-1/a+(r_e/2)p^2-ip} \ ,
\label{MW}
\end{eqnarray}
as long as $p \ll 1/r_e$.  
Therefore one expects that, at least for
small kinetic energies, $NN$ elastic scattering and meson production
in $NN$ collisions with a $NN$ final state exhibit the same energy dependence \cite{book,migdal},
which indeed was experimentally confirmed.
This treatment of FSI effects is often referred to as Migdal-Watson (MW) approach \cite{migdal}. 

In order to examine the reliability of the three methods described above we took different
$YN$ models from the literature \cite{Holz,Melni2,NijV,Haiden} and calculated the 
production amplitude $A(m^2)$ utilizing the meson exchange model from Ref. \cite{model}.
Then this amplitude was used for 
extracting the scattering length by means of the dispersion integral
Eq. (\ref{final}) or from the approximative prescriptions given by
Eqs. (\ref{arform}) and (\ref{MW}). For comparison we considered also 
the $^1S_0$ partial wave of the $np$ system of the Argonne potential \cite{Argonne}.
In this case $A(m^2)$ was set equal to the scattering wave function $\Psi (p,r)$
at the origin, more precisely to $\Psi^-(p,0)^*$, which corresponds to the assumption 
that the production operator is point-like. 

\begin{table}[h]
\caption{$S$-wave scattering lengths $a$ (in fm) for various $YN$ \cite{NijV,Melni2}
and $NN$ \cite{Argonne} potentials. 
The results for the original models are compared with those obtained by
applying the dispersion integral method (\ref{final}) 
and the approximations Eq. (\ref{arform}) (Jost-ERA) and Eq. (\ref{MW}) (MW).
}
\vskip 0.2cm
\begin{tabular}[t]{|c|c|c|c|c|}
\hline
model& \ exact result \ & \ disp.int. \ & \ Jost-ERA \ & \ MW \ \\
\hline
\hline
J\"ulich 01 singlet&    -1.02&-1.03&-1.28& -1.67\\
\hline
Nijmegen 97a singlet&   -0.73&-0.75&-0.98&-1.33 \\
\hline
Nijmegen 97f singlet&   -2.59&-2.57&-2.96& -3.35\\
\hline
J\"ulich 01 triplet&    -1.89&-1.66&-2.05& -2.42\\
\hline
Nijmegen 97a triplet&   -2.13&-1.98&-2.37& -2.75\\
\hline
Nijmegen 97f triplet&   -1.69&-1.61&-2.00& -2.37\\
\hline
Argonne v14 singlet &   -23.71&-23.54&-24.56& -24.79\\
\hline
\end{tabular}
\label{Table1}
\end{table}

Some selective results 
(for the Nijmegen NSC97 \cite{NijV} and J\"ulich 01 \cite{Melni2} $YN$ models 
and the Argonne v14 \cite{Argonne} $NN$ potential) are summarized in Table \ref{Table1}. 
The second column contains the 
correct scattering length evaluated directly from the potential model. 
One can see that the extraction of the scattering length via the 
dispersion integral (\ref{final}) yields results pretty close to the
original values for all considered potentials. In fact, in most cases
the deviation is significantly smaller than the uncertainty of the method, 
estimated in Ref. \cite{Gasparyan2003} to be 0.3 fm. The results of the 
Jost-ERA approach Eq. (\ref{arform}) 
exhibit a systematic offset in the order of 0.3 fm. The situation is much
worse for the Migdal-Watson approach Eq. (\ref{MW})
where a similar offset is found though
now in the order of 0.6 fm. As a consequence, the extracted values 
differ by 50 \%  or more from the correct scattering lengths. Only for
the $^1S_0$ $np$ partial wave the disagreement is still in the order of 5 \%.  
Here the reliability of the Jost-ERA and Migdal-Watson approaches are
comparable. This is in agreement with the expectations mentioned above. 

\begin{figure}[h]
\begin{center}
\epsfig{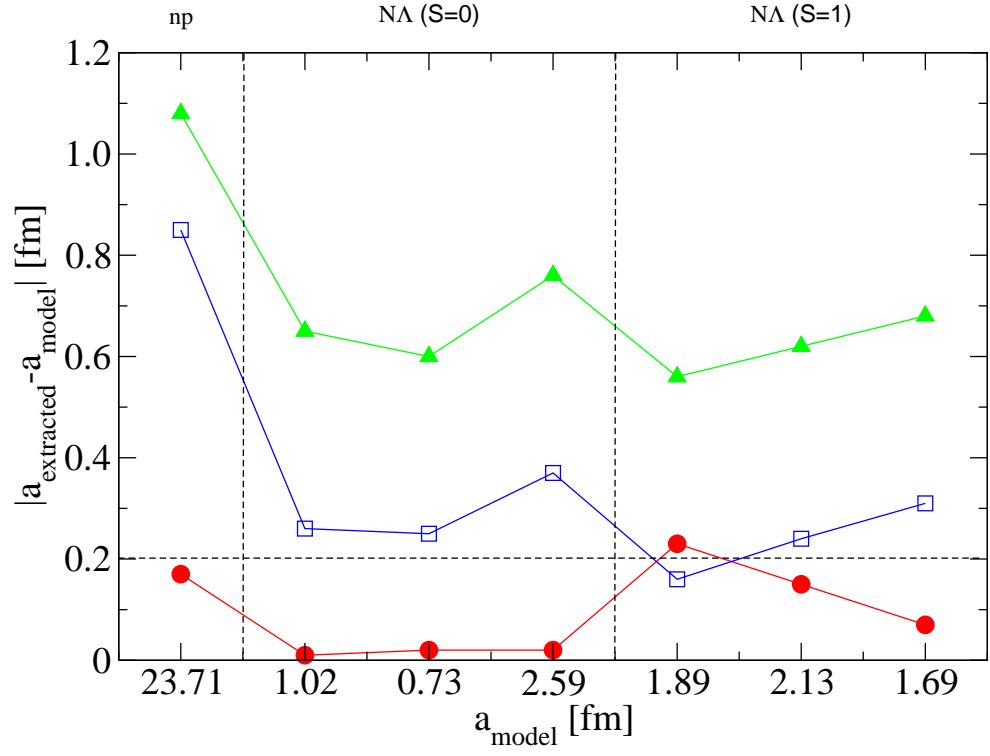}
\end{center}
\caption{(Color online) Comparison of different extraction methods for the
scattering length $a$. Shown are the differences between results 
predicted by various $YN$ and $NN$ models and 
corresponding values extracted via the dispersion integral method 
(circles), the Jost-ERA approach (\ref{arform}) (squares)
and the Migdal-Watson prescription (\ref{MW}) (triangles). 
The lines are drawn to guide the eye. 
}
\label{fig0}
\end{figure}

The systematic offset inherent in the Jost-ERA approach as well as in
the Migdal-Watson prescription can be best seen in Fig.~\ref{fig0}, where we
shown the difference between the scattering lengths predicted by various 
models and the values 
extracted via the dispersion integral (circles), the Jost-ERA method (squares)
and the Migdal-Watson prescription (triangles). 

While the Jost-ERA approach might still be a reasonable tool for getting a
first rough estimate of the scattering length for a particular two-body interaction
one should be rather cautious when using it for more quantitative analyses.  
In particular, its application in a combined fit to elastic scattering data and
invariant mass spectra, e.g. to $\Lambda p$ and $pp\to K^+\Lambda p$,  
is rather problematic and can easily cause misleading results. Because of the
offset in the scattering length in applications to final-state effects it is clear
that a combined fit cannot converge to a unique (the ``true'') $\Lambda p$
scattering length. Only the elastic data will favour values close to the 
``true'' scattering length whereas the production data tend to support larger
(negative) values. This is obvious from the corresponding Jost-ERA
results presented in Table \ref{Table1} and also from Fig.~\ref{fig0}. 
We believe that the analysis of Hinterberger and Sibirtsev presented 
in Ref. \cite{Hinter} is an instructive exemplification of this dilemma. 
Employing the 
Jost-ERA approach to low energy total $\Lambda p$ cross sections 
\cite{Alex,Sechi} and to experimental results for the missing mass 
spectrum of the reaction $pp\to K^+X$ \cite{Siebert} separately, they derived
(spin averaged) 
scattering lengths of $a=-1.81^{+0.18}_{-0.21}$ fm and $a=-2.57^{+0.20}_{-0.23}$ fm, 
respectively. Taking into account the error bars this is roughly the difference we 
would expect from the offset (of around $0.3$ fm) seen in our test calculations 
and, therefore, one must consider the results as being practically consistent 
with each other. 
But the authors of Ref. \cite{Hinter} attempted to ``reconcile'' the results
even more by introducing a spin-dependence in the fitting procedure.  
Indeed, 
with the relative magnitude of singlet to triplet contribution in the production
reaction as free parameter (their relative strength in the elastic channel is
fixed at 1:3 by the spin weight!) a ``fully consistent'' description of the 
combined data could be achieved \cite{Hinter} and apparently the spin-singlet as 
well as spin-triplet $\Lambda p$ $S$-wave scattering lengths could be 
determined from spin-averaged observables. 
Our experience with the Jost-ERA approach reported above, however,  
strongly suggests that the sensitivity to the spin seen in this analysis is most 
likely just an artifact of the method applied. 
 
\begin{table}[h]
\caption{$S$-wave effective ranges $r_e$ (in fm) for various $YN$ \cite{NijV,Melni2}
and $NN$ \cite{Argonne} potentials. 
The results for the original models are compared with those obtained by
applying the dispersion integral method (\ref{final}) and the approximation 
Eq. (\ref{arform}) (Jost-ERA).
}
\vskip 0.2cm
\begin{tabular}[t]{|c|c|c|c|}
\hline
model& \ exact result \ & \ disp.int. \ & \ Jost-ERA \ \\
\hline
\hline
J\"ulich 01 singlet&    4.49&4.31 &2.48\\
\hline
Nijmegen 97a singlet&   6.01&4.78 &2.81\\
\hline
Nijmegen 97f singlet&   3.05&2.82 &1.60\\
\hline
J\"ulich 01 triplet&    2.57& 2.49&1.89\\
\hline
Nijmegen 97a triplet&   2.74&2.60  &1.70\\
\hline
Nijmegen 97f triplet&   3.34& 2.67&1.69\\
\hline
Argonne v14 singlet&   2.78 & 2.91&0.43 \\
\hline
\end{tabular}
\label{Table2}
\end{table}

\begin{figure}[h]
\begin{center}
\epsfig{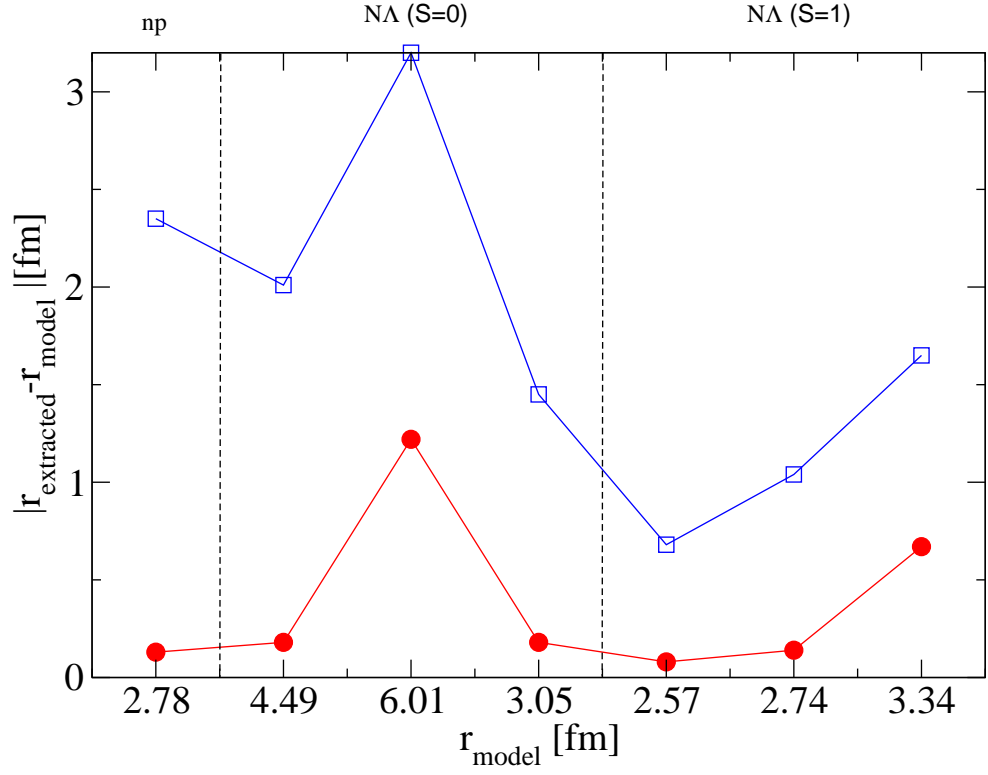}
\end{center}
\caption{(Color online) Comparison of different extraction methods for the
effective range $r_e$. Shown are the differences between results
predicted by various $YN$ and $NN$ models and
corresponding values extracted via the dispersion integral method
(circles) and the Jost-ERA approach (\ref{arform}) (squares).
The lines are drawn to guide the eye.
}
\label{fig0a}
\end{figure}

Let us now come to the effective range $r_e$. 
Since the dispersion relations yield only an integral representation for the product 
$a^2((2/3)a-r_e)$ but not for the effective range $r_e$ alone 
\cite{Gasparyan2003} it follows that the attainable accuracy of $r_e$ is always 
limited roughly by twice the relative error on $a$. 
Still, it is interesting to see what values one gets for $r_e$ from the dispersion integrals.
Corresponding results are presented in Table \ref{Table2} and in Fig. \ref{fig0a}. 
Evidently, the values extracted via the dispersion integral agree much better with the
original results as one might have expected. In fact, in practically all cases the 
deviation is in the order of only 5 \% or even less. This suggests that one could use
the dispersion integrals also to extract the effective range $r_e$ from data. But
one should keep in mind that, unlike the case of the scattering length, now one cannot
rely on a solid and general estimate of the uncertainty. 
As far as the Jost-ERA approach is concerned it is clear from Table \ref{Table2} that
it yields rather poor results. In case of the Migdal-Watson prescription (\ref{MW}) 
it turned out that the fit always prefers an effective range $r_e$ equal to 
zero. This is due to the term proportional to $r^2_ep^4$ in the denominator of the 
$A(m^2)$ that make the production cross section decrease too fast as compared 
to the data (or to our calculations with realistic models). 
Therefore we don't show any results of the Migdal-Watson fit for $r_e$.

One should note here that the upper limit in the dispersion integrals was
always taken such that $p_{max}=205$  MeV/c (as in
\cite{Gasparyan2003}) that corresponds to $\epsilon_{max}\equiv
m_{max}-m_0\approx 40$ MeV for the $\Lambda N$ and $\Sigma N$ systems and 
$\epsilon_{max}\approx 45$ MeV for the $NN$ system. In the latter case it
is interesting to see what happens if one varies the range of integration,
since the energy structure in the $NN$ interaction is much narrower due to
the large $NN$ scattering length. For $\epsilon_{max}=10$ MeV and 
$\epsilon_{max}=20$ MeV as upper limits of the integration one gets the
scattering lengths $a=-22.62$ fm and $a=-23.17$ fm, respectively -- which
are in principle still close to the original value. For the effective range,
however, the calculation yields $r_e=4.78$ fm and $3.72$ fm, respectively.
The reason why the agreement for $\epsilon_{max}>40$ MeV is so good
is that the $NN$ $^1S_0$ phase shift becomes sufficiently small at such 
energies, which implies a small uncertainty according to the error estimation 
in Ref.~\cite{Gasparyan2003}.

\section{Dispersion relation in the presence of Coulomb repulsion}

In the case when both baryons in the final state carry charges 
(for example in the reaction $pp\to K^0 p\Sigma^+$) there 
is a Coulomb interaction between them. Then the production amplitude
$A(m^2)$ acquires additional singularities at $p=0$, due to the 
long-range nature of the Coulomb forces, and the formalism developed 
in Sect. 2 is no longer applicable directly. In this section we want
to describe the modifications that are necessary in order to adapt the
dispersion-relation method to the situation when the Coulomb force
is present in the final-state interaction. We restrict ourselves to the
case of a repulsive Coulomb interaction so that no bound states are
present. 
 
In order to elucidate the principle idea we start
out from the case of elastic (two-body) scattering. Here the problem
can be most conveniently dealt with by applying the Gell-Mann--Goldberger
two-potential formalism \cite{Gell}. Let us assume that the total potential
$V = V_c + V_s$ is given by the sum of a short-ranged hadronic potential
$V_s$ and the Coulomb interaction $V_c$. Then the total reaction 
amplitude $T$ can be written as $T = T_c + T_{cs}$, where $T_c$ is the
Coulomb amplitude and $T_{cs}$ is defined by  

\be
T_{cs}=(1+T_cG_0)t_{cs}(1+G_0T_c), 
\ee
where $t_{cs}$ fulfils a Lippmann-Schwinger equation, 
\be
t_{cs}=V_s+V_sG_ct_{cs},
\ee
with the short-range potential $V_s$ as driving term. 
To obtain the physical on-shell amplitudes one needs to project
the corresponding $T$-operators on the so-called Coulombian asymptotic
states $\ket{p_\infty\pm}$ which are related to the Coulomb scattering states
(with fixed angular momentum -- in our case $l=0$) $\ket{p\pm}_c$ via
$ \ket{p\pm}_c=\ket{p_\infty\pm} + G_0^{\pm}T_c^{\pm}\ket{p_\infty\pm} $ \cite{vanHaeringen1976}.
Here $p$ denotes the center of mass momentum in the baryon-baryon system.
In this way one obtains in particular
\be
_c\bra{p-}t_{cs}\ket{p+}_c=-\frac{1}{\pi\mu}f_{cs}(p),
\ee
where $f_{cs}$ is the so-called Coulomb-modified nuclear scattering amplitude
and $\mu$ is the reduced mass. 
Its relation to the phase shift $\delta_{cs}$ is the following
\be
f_{cs}=\frac{e^{2i\delta_c}(e^{2i\delta_{cs}}-1)}{2ip},
\ee
with $\delta_c$ denoting the pure Coulomb $S$-wave phase shift 
given by \hbox{$\delta_{c}=\arg{(\Gamma(1+i\eta))}$} with 
$\eta=\frac{\mu e^2}{p}$. 

It has been shown in Ref.  \cite{Hamilton1973}
under rather general assumptions that 
the modified amplitude $\tilde{f}(p)=e^{-2i\delta_c}f_{cs}(p)/C^2(p)$
is free of the Coulomb singularities on the physical sheet
and possesses only the singularities caused by dynamical cuts 
(see also Refs. \cite{Cornille1962,vanHaeringen1977,Heller1966,Scotti1965}).
In addition below the inelastic cuts and above the two-baryon threshold
the modified unitarity relation reads
\be
\tilde{f}(s+i0)-\tilde{f}(s-i0)=2ip\tilde{f}(p)\tilde{f}^*(p)C^2(p)
\ee
with $ C^2(p)=\frac{2\pi\eta}{e^{2\pi\eta}-1}$ being the Coulomb
penetration factor. 
 
Furthermore an effective range function, modified for the presence of 
the Coulomb interaction, can be defined as well. Is is given by 
\be
S(p)\equiv pC^2(p)\cot\delta_{cs}(p)+Q(p)=-1/a_{cs}+r_ep^2/2+..,
\label{eff_range}
\ee
where $Q(p)\equiv\mu e^2[\psi(i\eta)+\psi(-i\eta)-2\ln\eta], \ \psi(z)=\Gamma'(z)/\Gamma(z)$.

Coming back now to the production reaction it can be analogously
shown that also   the modified production amplitude 
\be
\tilde{A}(m^2)=e^{-i\delta_c}A(m^2)/C(p)
\ee
is free of the aforementioned singularities \cite{Hamilton1973}.
Therefore a dispersion relation similar to Eq.~\eqref{dispers} can be written down
\be
\tilde A(m^2)=\frac1\pi\int_{-\infty}^{\tilde m\, ^2} \frac{\tilde D(m' \, ^2)}{m' \, ^2-m^2}dm' \, ^2
+\frac1\pi\int_{m_0^2}^\infty \frac{\tilde D(m' \, ^2)}{m' \, ^2-m^2}dm' \, ^2.
\label{dispers1}
\ee
Unitarity implies that the discontinuity for the elastic cut is 
\be
\tilde D(m^2) = 
\tilde A(m^2)e^{-i\delta_{cs}}\sin{\delta_{cs}}.
\ee
The solution to Eq.~\eqref{dispers1} is found in complete analogy to  
the case without the presence of the Coulomb interaction,
\be
\tilde A(m^2)=\exp\left[{\frac1\pi\int_{m_0^2}^{m_{max}^2}\frac{\delta_{cs}(m' \, ^2)}{m' \, ^2-m^2-i0}dm' \, ^2}\right]
\tilde \Psi(m^2) \label{C_intc} \ ,
\label{A_def1}
\ee
where $\tilde\Psi(m^2)$ is some function slowly varying with $m^2$.
If one neglects the the weak $m^2$ dependence present in $\tilde\Psi(m^2)$
the expression for the phase shift $\delta_{cs}$ in terms of the 
differential cross section becomes
%\be
%\nonumber
%&&\frac{\delta_{cs}(m^2)}{\sqrt{m^2-m_0^2}}\approx \\
%& & -\frac1{2\pi}{\bf P}
%\int_{m_0^2}^{m_{max}^2}
%\frac{\log{\left[\displaystyle{\frac{1}{p'C^2(p')}\frac{d^2\sigma}{dm' \, ^2dt}}\right]}}
%{\sqrt{m' \, ^2-m_0^2} \ (m' \, ^2-m^2)}\sqrt{\frac{m_{max}^2-m^2}{m_{max}^2-m' \, ^2}}dm' \, ^2.
%\label{finalc}
%\ee
\be
\frac{\delta_{cs}(m^2)}{\sqrt{m^2-m_0^2}} = 
-\frac1{2\pi}{\bf P}
\int_{m_0^2}^{m_{max}^2}
\frac{\log{\left[\displaystyle{\frac{1}{p'C^2(p')}\frac{d^2\sigma}{dm' \, ^2dt}}\right]}}
{\sqrt{m' \, ^2-m_0^2} \ (m' \, ^2-m^2)}\sqrt{\frac{m_{max}^2-m^2}{m_{max}^2-m' \, ^2}}dm' \, ^2.
\label{finalc}
\ee
Using the effective range expansion (\ref{eff_range}) one can then extract the
scattering length $a_{cs}$ from this dispersion integral. 

\section{Test of the method for the Coulomb case}

One of the obvious reactions for applying the formalism with
Coulomb is $pp\to K^0\Sigma^+p$ where one could determine the
$\Sigma N$ scattering length for the isospin 3/2 state. 
Note that the $\Sigma^+p$ channel does not couple to the $\Lambda N$
system and is therefore free of inelastic cuts (that start already on 
the left-hand side) as 
required for the applicability of the disperson integral method.  
For this reaction one could perform a model calculation analogous to 
the one for $pp\to K^+\Lambda p$ \cite{model} which we used for 
testing the dispersion-integral method in the absence of the
Coulomb interaction \cite{Gasparyan2003}. 
However, the implementation
of Coulomb effects into our momentum-space code is technically 
complicated and requires also some approximations \cite{Hanhart}. 
Thus, for the present test calculation we adopt a different
strategy. First, instead of the momentum-space $YN$ models
of Refs. \cite{Holz,Melni2,Haiden} we take the r-space Argonne 
($NN$) potential, however, with parameters modified in such a way 
that the effective range parameters are similar to those 
predicted by realistic $YN$ potentials \cite{Holz,NijV,Melni2,Haiden} 
for the $\Sigma N$ $I=3/2$ $^1S_0$ partial wave. 
In particular we prepared two models with Coulomb modified 
scattering length of $a_{cs}=-3.24$ fm (model 1) and 
$a_{cs}=-1.86$ fm (model 2), respectively. The corresponding scattering
lengths without 
Coulomb interaction are $-4.11$ fm  and $-2.01$ fm, respectively. 
For the transition amplitude we use the scattering wave function
calculated from those potential models and evaluated at the origin.
This corresponds to the assumption that the production operator is
point-like, which is reasonable as long as we are interested only in the 
mass dependence of the production amplitude. The corrections
stemming from a possible mass dependence of the production 
operator were discussed in Ref.~\cite{Gasparyan2003}.

\begin{figure}[h]
\begin{center}
\epsfig{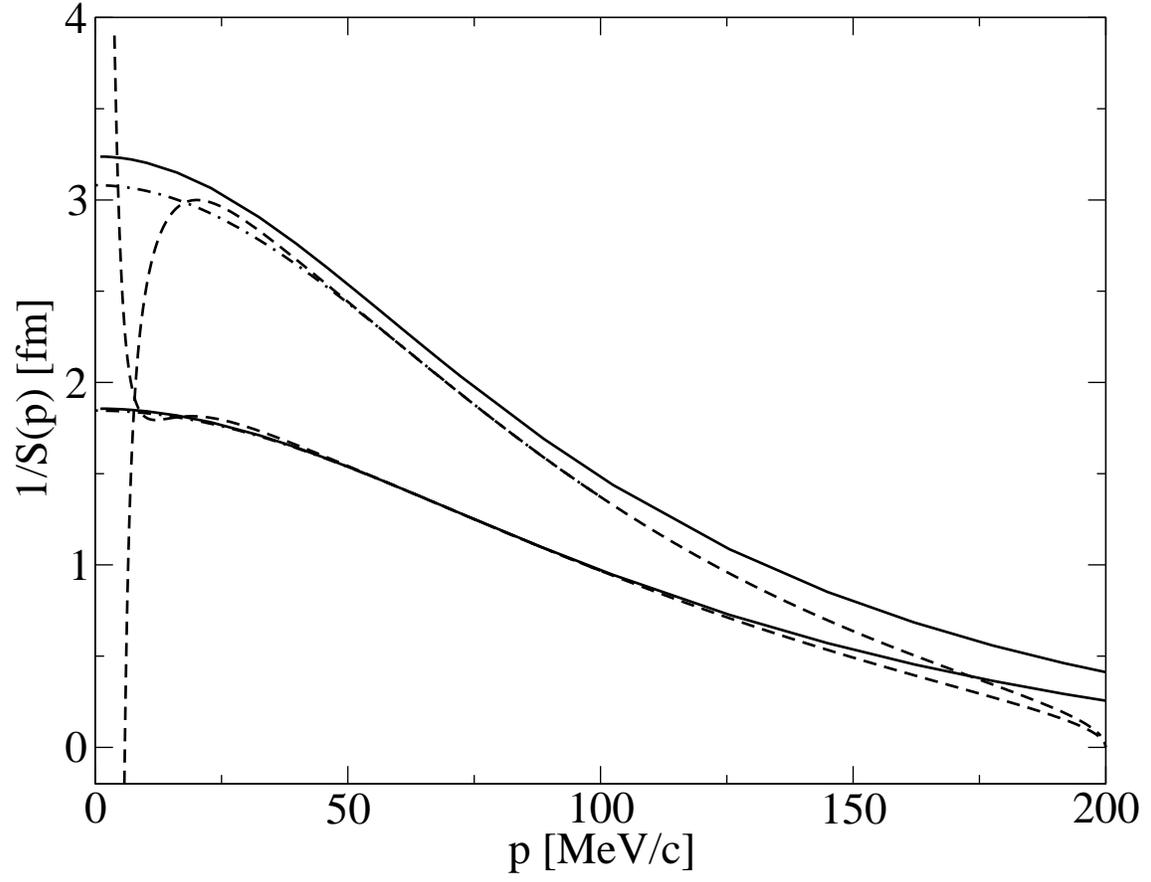}
\end{center}
\caption{The inverse of the effective range function $S(p)$
for model 1 (upper curves) and model 2 (lower curves). The 
solid lines denote the phase shifts predicted by the corresponding
model, whereas the dashed lines correspond to the phase shift extracted
via Eq.~\eqref{finalc}. The dash-dotted lines show the result of 
the smooth extrapolation of the dashed lines, as explained in the text.}
\label{fig1}
\end{figure}

The results of applying Eq.~\eqref{final} with $m_{max}-m_0=40$ MeV
($p_{max}=205$ MeV/c)
are shown in Fig.~\ref{fig1}, where we plot the function 
$1/S(p)$ which should coincide with the scattering length $-a_{cs}$
at $p=0$. Obviously, there is a strongly nonanalytic behavior of the 
extracted inverse effective range function when approaching the
threshold -- which, however, can be easily understood.
It is clear from Eq.~\eqref{eff_range} that the threshold 
behavior of the Coulomb modified phase shift is 
$\delta_{cs}\approx -a_{cs}\,p\,C^2(p)$, i.e. $\delta_{cs}$
goes to zero very rapidly. To obtain such a 
behavior on the left-hand side of Eq.~\eqref{finalc}
one needs to have a very precise cancellation in the integral
on the right-hand side of Eq.~\eqref{finalc} that is,
of course, impossible if one truncates the integral 
at a finite momentum. But still one can expect
for Eq.~\eqref{finalc} to work for momenta not too close to the threshold, 
namely above the typical Coulomb scale of $2\pi\alpha/\mu\approx 25$ MeV/c
where the factors $C^2(p)$ and $Q(p)$ that appear in the effective range
function (Eq. (\ref{eff_range})) become smoother. 
This is indeed the case, as can be seen from 
Fig.~\ref{fig1}. Thus, a natural and practical step here would be to
extrapolate the extracted $S(p)$ to the threshold from above.
Using a 4-th order polynomial of the type $-1/a_{cs}+r_ep^2/2-Pr_e^3p^4$ 
and fixing the coefficients 
in the region $50-100$ MeV/c, i.e. well above the Coulomb structure, 
one can then extrapolate $S(p)$ to the threshold, 
cf. the dash-dotted lines in Fig.~\ref{fig1}. 
In this case one gets a satisfactory agreement between the true and extracted 
scattering lengths. In fact, the deviations are not worse
then in the case when we consider the same potentials without 
Coulomb interaction and they are also within the theoretical error of 
0.3 fm estimated in Ref. \cite{Gasparyan2003}. The extracted values are 
$a_{cs}=-3.10$ fm and $a_{cs}=-1.86$ fm for models 1 and 2,
respectively, with Coulomb, and $a_s=-4.05$ fm and $a_s=-2.06$ fm
when the Coulomb interaction is switched off. We also checked that
the sensitivity of the result to the region of interpolation of the
effecitve range function $S(p)$ is rather low. For instance if one shifts
the lower bound of this region to $70$ MeV/c the corresponding change
in the scattering length will be less then $0.05$ fm.

\begin{figure}[h]
\begin{center}
\epsfig{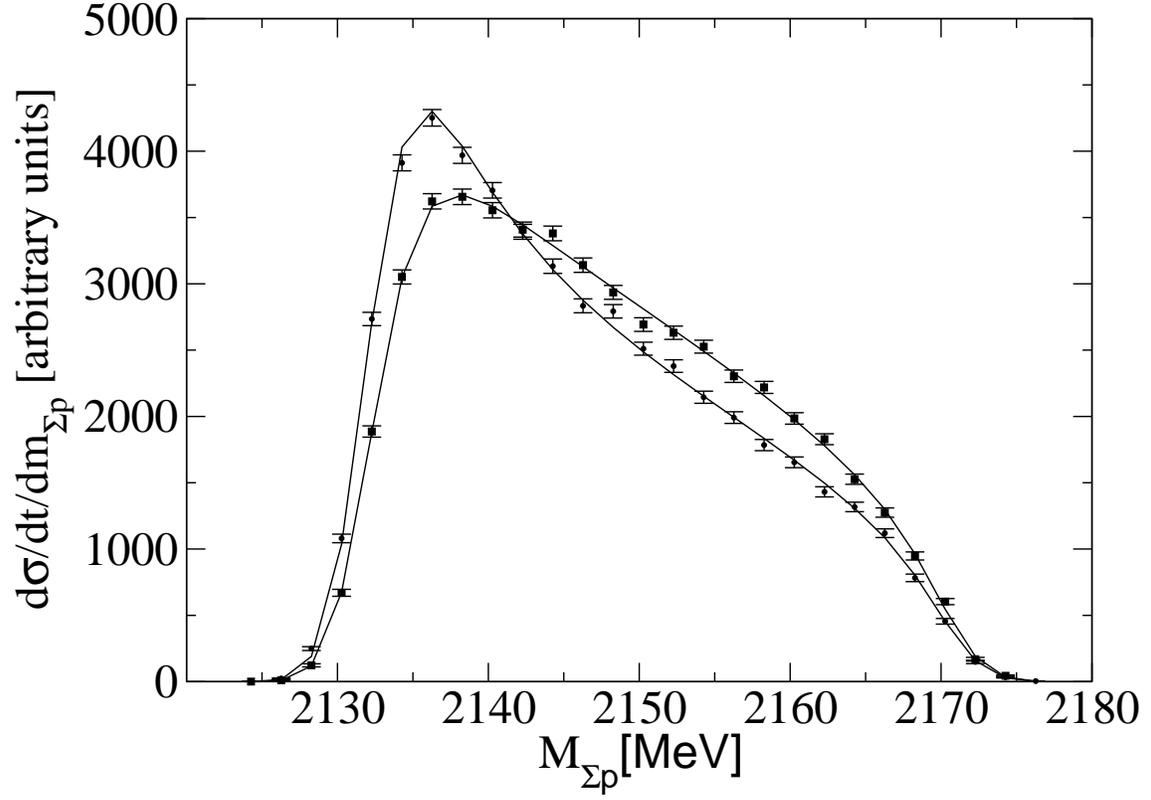}
\end{center}
\caption{Pseudo data for the 
differential $\Sigma^+p$ cross section generated from
model 1 (circles) and model 2 (squares) as a function of
the $\Sigma p$ invariant mass $M_{\Sigma p}$ with corresponding 
fit by the exponential parameterization Eq. (\ref{fit1}). 
}
\label{fig2}
\end{figure}

If one wishes to consider a more realistic
situation, one needs to deal with mass distributions with
finite statistical errors, finite mass resolution,
and finite binning. To examine also this situation we have generated
two data sets, corresponding to the models 1 and 2
as shown on Fig.~\ref{fig2}. We have chosen the 
binning as well as the mass resolution to be equal to $2$ MeV
(the same as in the experiment \cite{Siebert} that was analyzed in \cite{Gasparyan2003}),
and the statistics to be rather high in order to minimize the influence of 
the statistical error bars on the results. The excess energy was set to $40$
MeV to simplify the simulation. In a realistic situation larger values are
preferable to minimize the influence of the meson-baryon interactions, 
cf. the corresponding remarks in Ref. \cite{Gasparyan2003}. 
In our test calculation such meson-baryon interactions are neglected
anyway. 

\begin{figure}[h]
\begin{center}
\epsfig{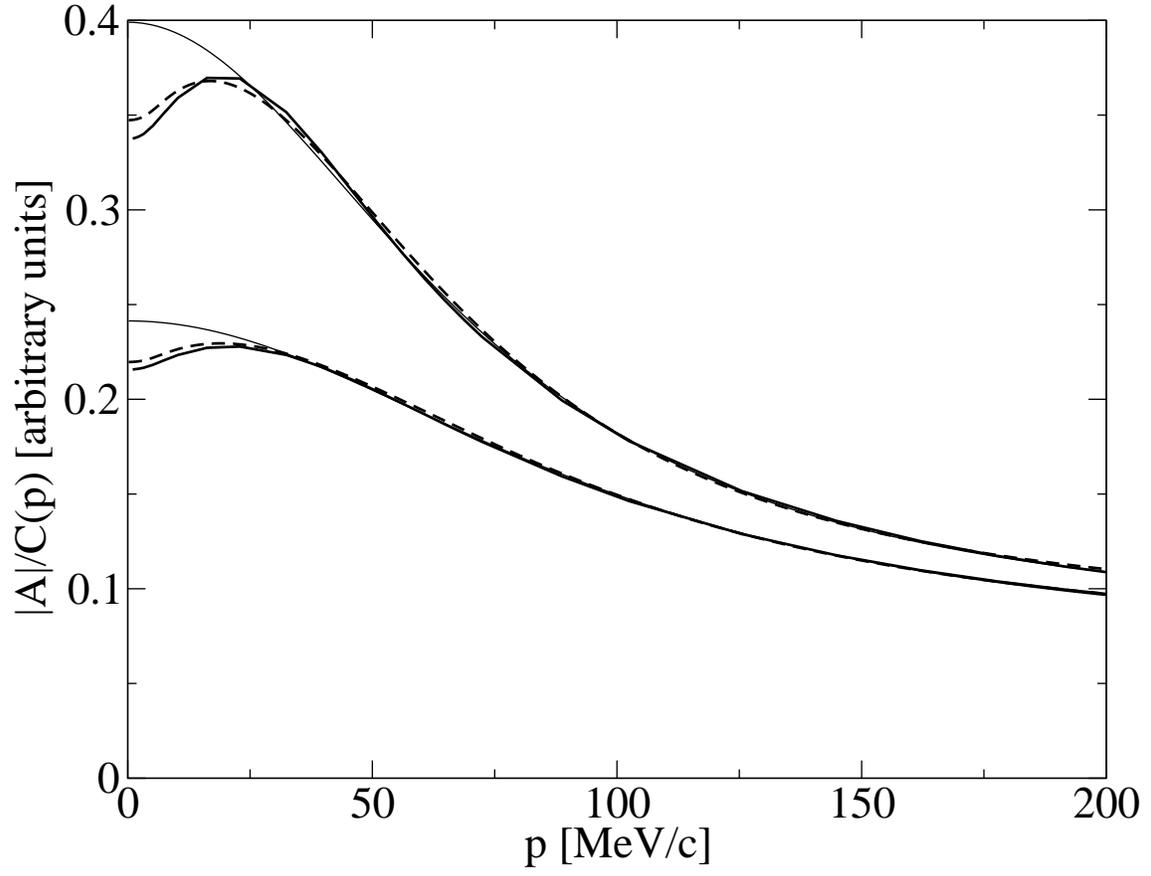}
\end{center}
\caption{The production amplitude $A(m^2)$ divided by the factor $C(p)$
for model 1 (upper curves) and model 2 (lower curves).
The solid lines are the amplitudes as calculated from the models. The dash-dotted lines
correspond to the fitted amplitudes. The dashed lines denote the fitted
amplitudes improved by the iterative procedure as discussed in the text.}
\label{fig3}
\end{figure}

We start here with the procedure suggested in appendix A of \cite{Gasparyan2003},
namely by fitting the cross section with an exponential parameterization of the type
\be
\frac{d^2\sigma}{dm \, ^2dt}=C^2(p) \times \exp{\left[C_0+\frac{C_1^2}{(m^2-C_2^2)}  \right]}
\times \text{phase space} \ . 
\label{fit1}
\ee
This formula fits the generated cross section with 
the $\chi^2$ per degree of freedom of $\chi^2_{dof}\sim 1$, cf. Fig.~\ref{fig2}.
A new problem that arises here is that the 
production amplitude contains a very narrow structure
(of the size $2\pi\alpha/\mu$) close to threshold as can
be seen in Fig.~\ref{fig3} (solid lines).
Clearly, this structure can not be reproduced after the fitting
procedure as it gets smeared out by the mass resolution and
binning (their size is much larger than the scale of the structure).
The amplitudes coming from the fit are depicted in Fig.~\ref{fig3}
by dash-dotted lines.
However the fit can be improved if one notes that
the structure comes mostly from the part of the dispersion integral
\eqref{A_def1} containing the leading term in the 
$\delta_{cs}$ expansion near the threshold, namely (in the nonrelativistic case)
\be
\exp\left[{\frac1\pi\int_{0}^\infty\frac{-a_{cs}C^2(p')}{p' \, ^2-p^2-i0}\left(\frac{p^2}{p'^2}\right)dp' \, ^2}\right]=\exp\left[-a_{cs}Q(p)\right],
\ee
where we made a subtraction at $p=0$ to render the integral convergent.
This does not change the energy dependence of the resulting exponent.
Indeed the structure disappears after dividing the production amplitude
by the factor $\exp[-a_{cs}Q(p)]$.

\begin{figure}[h]
\begin{center}
\epsfig{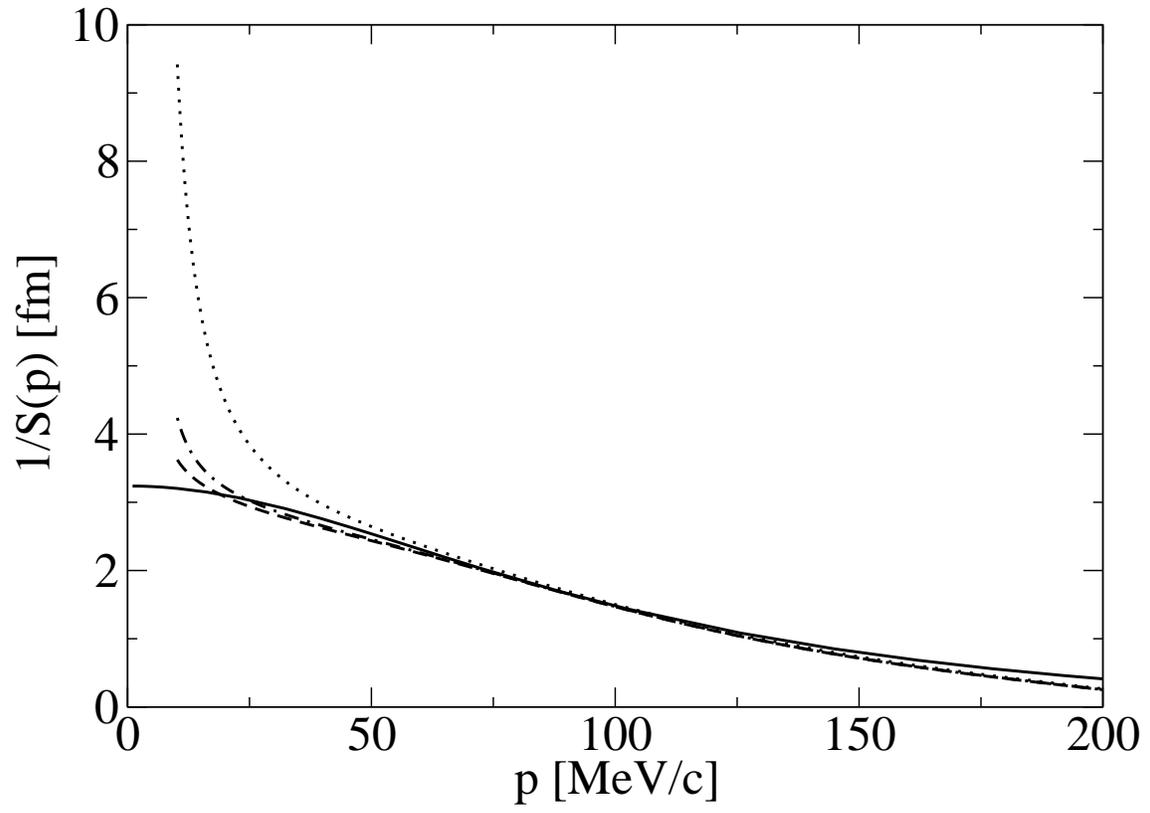}
\end{center}
\caption{The inverse of the effective range function $S(p)$ calculated by means of
the iterative procedure as discussed in the text. The dotted line corresponds
to the zeroth iteration, the dashed line corresponds to the first
iteration, and the dash-dotted line corresponds to the second iteration. The
solid curve denotes the exact result. Shown are results for the
model 1.
}
\label{fig4}
\end{figure}

Obviously the scattering length is unknown before its
extraction! But what can be done is to resort to an iterative procedure
including first the extraction of the unimproved scattering length,
then putting it into the fit function,
\be
\frac{d^2\sigma}{dm \, ^2dt}=C^2(p)\times\exp{\left[C_0+\frac{C_1^2}{(m^2-C_2^2)}
-2a_{cs}Q(p)  \right]} \times \text{phase space} \ ,
\label{fit2}
\ee
and repeating this step until the procedure converges. Fortunately 
the iterations converge already after three or four iterations, 
and the resulting amplitudes are shown in Fig.~\ref{fig3}
by the dashed lines. The improvement of the fit is quite 
obvious.
Finally we applied the combination of the extrapolation and iteration
procedures to  obtain the scattering lengths from the pseudodata.
Since the data have a statistical uncertainty we generated a sample of
1000 mass distributions and looked at the average value of the scattering 
lengths. They turned out to be $-2.89\pm 0.06$ fm for model 1 and  $-1.82\pm
0.05$ for model 2. Note that the deviation from the correct values is now a 
bit larger, but still reasonable ($~0.35$ fm in the worst case).
Fig.~\ref{fig4} shows how the extracted inverse effective range function 
approaches to the correct one given by the model by the example of model 1.

\section{Summary}

In a recent publication \cite{Gasparyan2003} we have presented a formalism
based on dispersion relations 
that allows one to relate spectra from large-momentum transfer reactions,
such as $pp\to K^+p\Lambda$ or $\gamma d\to K^+n\Lambda$, directly to
the scattering length of the interaction of the final-state particles.
An estimation of the systematic uncertainties of that method, relying on general
arguments, led to the conclusion that the theoretical error in the extracted
scattering length should be less than 0.3 fm. This finding was corroborated 
in an application of the method to results of a microscopic model calculation 
for $pp\to K^+p\Lambda$. 

In the present paper this dispersion theoretical method was generalized to
the case where a repulsive Coulomb force is present in the final-state interaction.
As an example let us mention the reaction $pp\to K^0p\Sigma^+$ which could
be used to extract the $p\Sigma^+$ scattering length.  
Though the generalization of the formalism itself is 
straight forward it turned out that there are some additional features
due to the Coulomb interaction which need to be taken into account 
in concrete applications of the method to data. 
These practical aspects were thoroughly discussed and it was shown how to
circumvent the difficulties. In a test calculation utilizing  
potential models with effective range parameters similar to those
of realistic $YN$ interactions the extracted values for the scattering lengths 
were found to agree within 0.3 fm with those predicted by the models.
Thus, the accuracy of the dispersion theoretical method for extracting the
scattering lengths from final-state interactions with Coulomb force is
comparable to the case where no Coulomb interaction is present. 

We presented also a more detailed examination of the accuracy of the
dispersion-integral method than in Ref. \cite{Gasparyan2003}. In particular we considered
final-state interactions of varying strengths, corresponding to a much larger range
of values of the scattering length. These investigations confirmed the reliability
of the general error estimate provided in Ref. \cite{Gasparyan2003}. Indeed in 
most of the considered cases the deviation of the extracted scattering length 
from the true value was significantly smaller than the uncertainty of 0.3 fm 
derived in that paper. 
In addition we studied the effective range $r_e$ which can be also extracted 
by the proposed dispersion-integral method. For most of the interaction models 
considered the extracted values of $r_e$ agreed remarkably good with the 
true results. Thus, it might be sensible to use the proposed method to extract
the effective range from data - though one should always keep in mind that
for this quantity a generally valid error estimation is not possible
\cite{Gasparyan2003}. 
 
Finally, we compared the present method with the performance of other, 
approximative treatments
of the final-state interaction that are commonly used in the literature to
extract information on the scattering length and also the effective range.
In particular we tested the Jost-function approach based on the effective-range
approximation (Jost-ERA) and an even simpler approach that relies simply on utilizing 
the effective range approximation itself. 
Thereby, we showed that the latter methods lead to systematic deviations 
from the true values of the scattering lengths in the order of 0.3 fm (Jost-ERA) 
and even 0.7 fm (direct effective-range approximation). 
This suggests that one should be rather cautious in the interpretation of results 
achieved with those methods.

\acknowledgments{
A.G. would like to acknowledge finanical support by the
grant No. 436 RUS 17/75/04 of the Deutsche Forschungsgemeinschaft.
Furthermore he thanks the Institut f\"ur Kernphysik at the 
Forschungszentrum J\"ulich for its hospitality during the period when
the present work was carried out.
}

\newpage

\newpage

\newpage

\newpage

\newpage

\newpage

\end{document}